\documentclass{amsart}\usepackage[]{graphicx}\usepackage[]{color}
\makeatletter
\def\maxwidth{ %
  \ifdim\Gin@nat@width>\linewidth
    \linewidth
  \else
    \Gin@nat@width
  \fi
}
\makeatother

\definecolor{fgcolor}{rgb}{0.345, 0.345, 0.345}

\usepackage{framed}
\makeatletter
 {\par\unskip\endMakeFramed%
 \at@end@of@kframe}
\makeatother

\definecolor{shadecolor}{rgb}{.97, .97, .97}
\definecolor{messagecolor}{rgb}{0, 0, 0}
\definecolor{warningcolor}{rgb}{1, 0, 1}
\definecolor{errorcolor}{rgb}{1, 0, 0}
\usepackage{alltt}

\usepackage{amsmath}
\usepackage{amssymb}
\usepackage{booktabs}
\usepackage{graphicx}
\usepackage{graphics}
\usepackage{setspace}
\usepackage{multicol}
\usepackage{multirow}
\usepackage{enumitem}
\usepackage{listings}
\usepackage{courier}
\usepackage{lipsum}
\usepackage{bm}
\usepackage{setspace}
\usepackage{adjustbox}
\usepackage{mathrsfs}
\usepackage{tabularx}
\usepackage{natbib}
\usepackage{subcaption}
\usepackage[titletoc,title]{appendix}
\usepackage[a4paper, total={6in, 8in}]{geometry}
\usepackage{moreverb}

\newcommand\BibTeX{{\rmfamily B\kern-.05em \textsc{i\kern-.025em b}\kern-.08em
T\kern-.1667em\lower.7ex\hbox{E}\kern-.125emX}}

\author{John Kidd}
\address{Department of Biostatistics, University of North Carolina at Chapel Hill,
Chapel Hill, North Carolina, U.S.A.}
\email{johnkidd@unc.edu}

\author{Dan-Yu Lin}
\address{Department of Biostatistics, University of North Carolina at Chapel Hill,
Chapel Hill, North Carolina, U.S.A.}
\email{lin@bios.unc.edu}

\title{Improving the Power to Detect Indirect Effects in Mediation Analysis}
\IfFileExists{upquote.sty}{\usepackage{upquote}}{}
\begin{document}

\thispagestyle{plain}

\begin{abstract}
Causal mediation analysis seeks to determine whether an independent variable affects a 
response variable directly or whether it does so indirectly, by way of a mediator. 
The existing statistical tests to determine the existence of an indirect effect are
overly conservative or have inflated type I error. In this article, we propose 
two methods based on the principle of intersection-union tests that offer improvements 
in power while controlling the type I error. We demonstrate the advantages of the proposed methods
through extensive simulation. Finally, we provide an application to a large proteomic study.

\end{abstract}

%%%% 

\maketitle

\section{Introduction}
\label{s_intro}
Causal mediation analysis seeks to determine the pathways by which an independent 
variable affects a response variable: either directly or through some additional variable. 
If the independent variable affects the response through a secondary variable, called 
a mediator, there is said to be an indirect effect \cite{Alwin1975}. 
Mediation analysis has been used extensively, especially in social sciences \cite{MacKinnon2007a}
and public health sciences \cite{Bellavia2019, Richiardi2013}, and has become increasingly popular in genetics 
\cite{Barfield2017, Cardenas2019, Huang2014, Hutton2018, Raulerson2019}.

Detecting an indirect effect is seemingly simple but actually very difficult 
\cite{Biesanz2010, Huang2016, Vanderweele2009, Zhong2019}. 
Let $\beta$ and $\gamma$ denote, respectively, the effect of the independent variable on the mediator
and the effect of the mediator on the response variable,
and let $\widehat \beta$ and $\widehat \gamma$ denote the corresponding maximum likelihood estimators, 
which are independent under a no unmeasured confounding assumption \cite{Imai2010}.                   
In the product of coefficients method \cite{Alwin1975, MacKinnon2007a}, 
the null hypothesis of no indirect effect   
means $\beta \gamma =0$, which is true if one of the two
parameters is zero or both are zero. 
If $\beta = \gamma=0$, the asymptotic distribution of $\widehat \beta \widehat \gamma$ 
is the product of two zero-mean normal random variables, rather than the normal distribution.
If either $\beta$ or $\gamma$ is non-zero, but not both, the asymptotic distribution
of $\widehat \beta \widehat \gamma$ a zero-mean normal \cite{Kisbu-Sakarya2014, Wang2018}. % change
In practice, one does not know which distribution is correct because both scenarios constitute the null hypothesis.
The well-known test of Sobel \cite{SobelME1982} uses the normal distribution for $\widehat \beta \widehat \gamma$
and is overly conservative if $\beta=\gamma=0$,
whereas the test based on the product-normal distribution \cite{MacKinnon2002, MacKinnon2004} is too liberal
if $\beta$ or $\gamma$ is non-zero. 
 
In this paper, we develop new methods to detect indirect effects based on the principle
of intersection-union tests \cite{Berger1997}. 
The proposed tests have correct type I error whether one or both parameters are zero
and have good power when both parameters are non-zero.
We demonstrate the advantages of the proposed methods 
through extensive simulation studies
and provide an application to 
the Sub-Populations and Intermediate Outcome Measures in 
COPD Study (SPIROMICS) \cite{Couper2014}. 

\section{Methods}
\label{s_methods}

\begin{figure}[ht]
\begin{center}
  \includegraphics[width=5in]{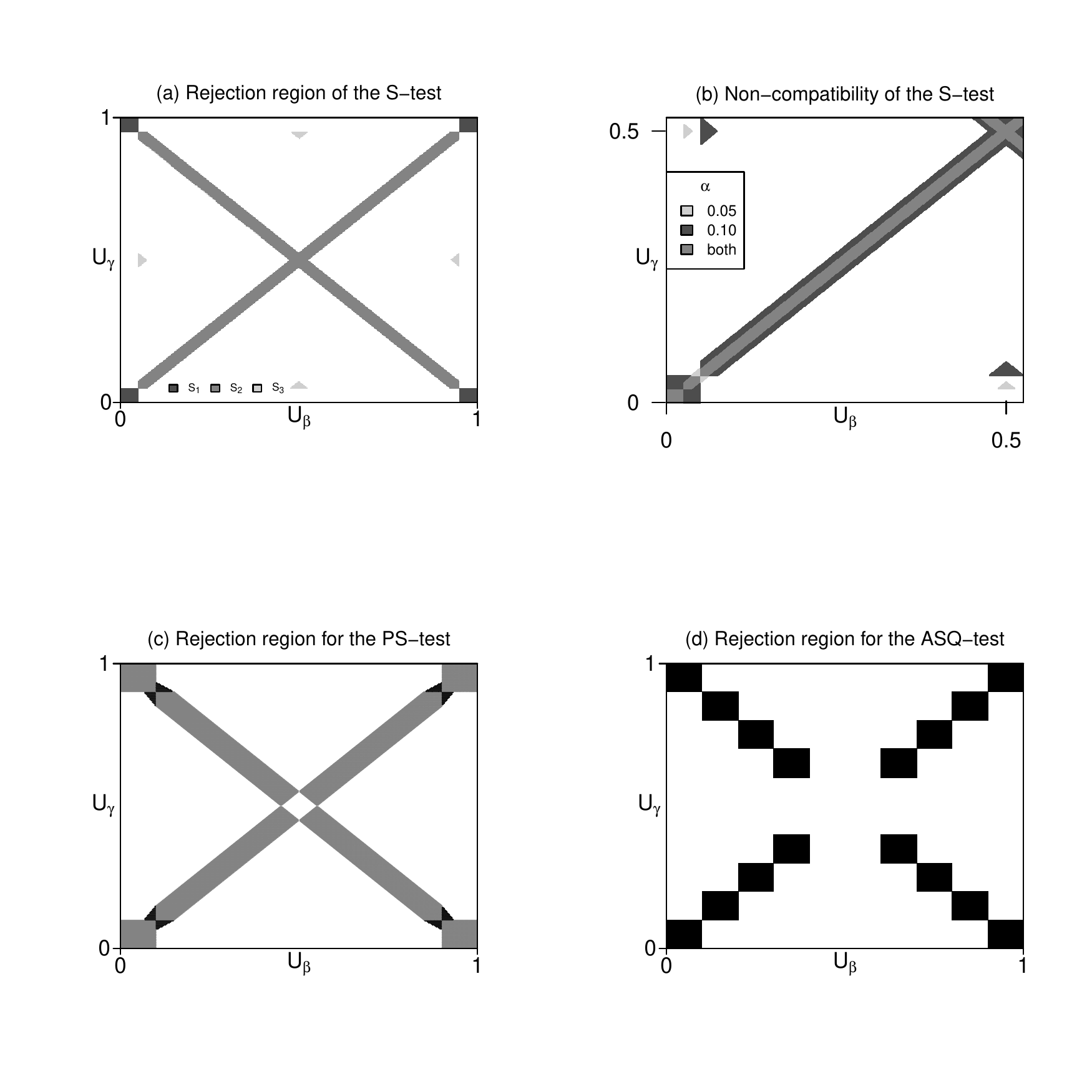}
\end{center}
   \caption{Rejection regions of the original S-test, the $p$-value approach S-test (PS-test),       
   and the ascending squares test (ASQ-test) using cumulative probabilities $U_\beta$ and $U_\gamma$:
   (a) the rejection region of the S-test as the union of $S_1$, $S_2$, and $S_3$;
   (b) non-compatibility of the S-test shown by the middle shade of gray, where the S-test
   rejects for $\alpha = 0.05$ only;
   (c) rejection region for the PS-test where portions of the original S-test are removed
   and the region $S_m$ (denoted by the darker shade) is added;
   (d) rejection region for the ASQ-test where the squares are constructed using the dimensions 
   of $S_1$ from the original S-test with multiple squares aligning on diagonal corners and 
   extending towards the center of the region.} 
\label{figComb}
\end{figure}

We are testing the null hypothesis $H_0:\{\beta=0\} \cup \{\gamma=0\}$ against the alternative
hypothesis $H_A:\{\beta\ne 0\} \cap \{\gamma \ne 0\}$. Thus, this problem can be cast within
the framework of intersection-union tests \cite{MacKinnon2002}. Berger \cite{Berger1997} 
proposed the so-called S-test 
for this problem with normally distributed data, although not in the context of mediation analysis.         
Let $T_{\beta}$ and $T_{\gamma}$ be the test statistics for testing the null hypotheses that             
$\beta=0$ and $\gamma=0$, respectively. Assume the statistics $T_{\beta}$ and $T_{\gamma}$ are independent
with distribution functions $F_{\beta}(\cdot)$ and $F_{\gamma}(\cdot)$, respectively.                
Write $U_{\beta} = F_{\beta}(T_{\beta})$ and $U_{\gamma} = F_{\gamma}(T_{\gamma})$.                  
We reject $H_0:\{\beta=0\} \cup \{\gamma=0\}$ if $(U_{\beta}, U_{\gamma})$ 
falls into the rejection region $S$ shown in Figure~\ref{figComb}a, which consists of three distinct 
regions, $S_1$, $S_2$, and $S_3$.
In $S_1$, $U_{\beta}$ and $U_{\gamma}$ are less than $\alpha/2$ or greater than $1 - \alpha/2$; 
in $S_2$, the difference between $U_\beta$ (or similarly $1 - U_\beta$) and $U_\gamma$ is less than 
$\alpha/4$;
and in $S_3$, $U_{\beta}$ and $U_{\gamma}$ are greater than $\alpha/2$ or less than $1 - \alpha/2$ 
and the difference between a specified $U$ value and 0.5 is small (or similarly, large) 
compared to the remaining $U$ value.
The S-test has level $\alpha$ because ${\rm Pr} \{ (U_{\beta}, U_{\gamma}) \in S\} = \alpha$
when either $U_{\beta}$ or $U_{\gamma}$ has the standard uniform distribution.

As shown in Figure~\ref{figComb}a, $S_1$ is comprised of the four squares in the corners.
In each square, $U_{\beta}$ and $U_{\gamma}$ are either less 
than $\alpha/2$ or greater than $(1 - \alpha/2)$. 
Thus, $(U_{\beta}, U_{\gamma}) \in S_1$ if $T_{\beta}$ and $T_{\gamma}$ are both
significant at the $\alpha$ level, i.e., the maximum of the two $p$-values is less than $\alpha$. 
We refer to the test based on $S_1$ alone \cite{cohen1983applied, MacKinnon2002} as maxP.    
Due to the additional rejection regions $S_2$ and $S_3$,                                          
the S-test is guaranteed to be more powerful than maxP.                          

Unfortunately, the S-test has some undesirable properties 
which make it inappropriate for testing $H_0$.                                  % change           
First, it can reject $H_0$ at a certain value of $\alpha$ but fail to do so at a
larger value of $\alpha$. An example of this non-compatibility is given in Figure~\ref{figComb}b:
$S_2$ for $\alpha = 0.05$ will reject $H_0$ for some values of ($U_\beta, U_\gamma$) 
that would not cause $H_0$ to be rejected for $\alpha = 0.10$ near the point where $S_1$ and $S_2$ meet;
the location of $S_3$ depends on the value of $\alpha$, with a smaller value of $\alpha$ causing $S_3$ 
to be closer to the edge of the graphed region, enclosing values of ($U_\beta, U_\gamma$) 
not included by any larger value of $\alpha$.
Second, the S-test may reject $H_0$ when the indirect effect is estimated at zero. 
Indeed, if $\hat\beta$ and $\hat\gamma = 0$, then $U_{\beta}=U_{\gamma}=0.5$,
which is the center of $S_2$.

We propose two methods to address the shortcomings of the S-test. 
In the first method, we alter the rejection region and use $p$-values                     
to determine significance; we will refer to this method as the PS-test.
Specifically, we remove $S_3$ since it is a major contributor to non-compatibility 
and is not a very meaningful rejection region. 
In addition, we remove the portion of $S_2$ near 
$(0.5, 0.5)$, where the two diagonal bands cross, so as to avoid rejecting $H_0$ 
when the indirect effect is estimated at zero. 
Finally, we define the $p$-value as the smallest value of $\alpha$ at which $H_0$ is rejected.   
This step enlarges the rejection region between $S_1$ and $S_2$,
as shown in Figure~\ref{figComb}c. The added
rejection region, denoted as $S_m$, takes the form of eight right triangles 
with base $\alpha/4$ and height $\alpha/6$. 

The addition of $S_m$ increases the type I error. 
If $\beta = \gamma = 0$, then both $U_{\beta}$ and $U_{\gamma}$ have the standard 
uniform distribution.
The total area of $S_m$ is $\alpha^2/6$. 
Removing $S_3$, which is comprised of four triangles each with base  
$\alpha/2$ and height $\alpha/4$, reduces the type I error by $\alpha^2/4$,
and the removal of the central section eliminates an additional $\alpha^2/8$. 
Thus, the final rejection probability is $\alpha \{1-(5\alpha/24)\}$,
such that the type I error rate is guaranteed not to be inflated.
If only one of the two parameters is equal to zero, the change
in the type I error is more complex. 
However, we show in the Appendix
that the potential inflation of the type I error is negligible. 

The second method we propose here also changes the rejection region of the original S-test 
and utilizes $p$-value thresholds. However, the rejection region is different from that of the PS-test
in order to avoid potential inflation of the type I error.
As shown in Figure~\ref{figComb}d, the rejection region consists of multiple squares 
of the size of $S_1$, the squares ascend 
(or descend) from each corner towards the center,
and each square meets at diagonal corners. 
We refer to this method as the ascending squares (ASQ) test. 

To preserve compatibility, the significance level thresholds must be chosen such that each level 
divides evenly into all larger levels. Without this restriction, the rejection region 
for a smaller significance level will fail to lie within that of larger levels.
To prevent inflation of the type I error, each level must divide evenly into $1.0$. 
This restriction ensures that the centermost squares will not have inappropriate overlapping.

Given the predetermined significance levels, the ASQ-test begins at the largest 
significance level and continues down to the next largest, determining at each level 
whether $(U_{\beta}, U_{\gamma})$ lies within the rejection region. If
$(U_{\beta}, U_{\gamma})$ is within the rejection region at a specific significance 
level, we conclude that the true $p$-value for the test is less than this significance 
level. By proceeding through the predetermined significance levels, the smallest 
significance value for which the null hypothesis $H_0$ is rejected is determined, and that 
value is considered the $p$-value threshold; the true $p$-value is less than this value. 

We may limit the degree that the bands for the PS-test or the squares for the ASQ-test are allowed 
to extend toward the center. This will alleviate the requirement that each $\alpha$ 
divides evenly into $1.0$ for the ASQ-test and also reduces the potential inflation of the type I error
by the PS-test. To avoid rejecting $H_0$ when $\hat\beta=\hat\gamma = 0$, 
the ASQ-test must omit the center-most squares. The decision on how far the bands or 
squares are allowed to extend is based on a trade-off between                                            
power and type I error.

\section{Simulation Studies}
\label{s_sim_results}

We conducted extensive simulation studies to evaluate the performance of the proposed 
and existing methods. We let independent variable $G$ be Bernoulli(0.5),
mediator $M = \beta  G + \epsilon_M$, and response variable 
$Y = 0.2 G + \gamma  M + \epsilon_Y$,  where $\epsilon_M$ and $\epsilon_Y$ 
are independent standard normal random variables. 
We varied $\beta$ and $\gamma$ from $0$ to $0.4$ and set the sample size
$n$ to 100, 500 or 1,000. 
For each combination of simulation parameters, we used 20,000 replicates
to estimate the type I error or power of each test at $\alpha=0.05$.
For the proposed methods, the band or squares were limited to 50\% 
of the possible extension. 

\setlength\tabcolsep{4pt}
\begin{table}[ht]
  \caption{Empirical Type I Error Rates at the Nominal Significance Level of 0.05}
  \label{t:typeI}
  \centering

\begin{tabular}{lll>{\raggedleft\arraybackslash}p{1.35cm}>{\raggedleft\arraybackslash}p{1.35cm}>{\raggedleft\arraybackslash}p{1.35cm}>{\raggedleft\arraybackslash}p{1.35cm}>{\raggedleft\arraybackslash}p{1.35cm}>{\raggedleft\arraybackslash}p{1.5cm}}
\toprule
\multirow{0}{*}[0pt]{$\beta$} & \multirow{0}{*}[0pt]{$\gamma$} & \multirow{0}{*}[0pt]{$n$} & \multirow{0}{*}[0pt]{Sobel} & \multirow{0}{*}[0pt]{maxP} & product normal & \multirow{0}{*}[0pt]{S-test} & \multirow{0}{*}[0pt]{PS-test} & \multirow{0}{*}[0pt]{ASQ-test}\tabularnewline
\midrule
0 & 0 & 100 & $<$0.001 & 0.003 & 0.054 & 0.053 & 0.026 & 0.025\tabularnewline
 &  & 500 & $<$0.001 & 0.003 & 0.048 & 0.049 & 0.025 & 0.024\tabularnewline
 &  & 1000 & $<$0.001 & 0.003 & 0.051 & 0.049 & 0.024 & 0.024\tabularnewline
\addlinespace
 & 0.1 & 100 & 0.001 & 0.008 & 0.110 & 0.049 & 0.033 & 0.032\tabularnewline
 &  & 500 & 0.005 & 0.028 & 0.303 & 0.050 & 0.047 & 0.047\tabularnewline
 &  & 1000 & 0.012 & 0.041 & 0.459 & 0.047 & 0.047 & 0.046\tabularnewline
\addlinespace
 & 0.2 & 100 & 0.004 & 0.023 & 0.261 & 0.047 & 0.042 & 0.042\tabularnewline
 &  & 500 & 0.029 & 0.051 & 0.613 & 0.051 & 0.051 & 0.051\tabularnewline
 &  & 1000 & 0.038 & 0.048 & 0.723 & 0.048 & 0.048 & 0.048\tabularnewline
\addlinespace
 & 0.3 & 100 & 0.013 & 0.042 & 0.435 & 0.051 & 0.051 & 0.050\tabularnewline
 &  & 500 & 0.041 & 0.052 & 0.740 & 0.052 & 0.052 & 0.052\tabularnewline
 &  & 1000 & 0.044 & 0.049 & 0.814 & 0.049 & 0.049 & 0.049\tabularnewline
\addlinespace
 & 0.4 & 100 & 0.023 & 0.048 & 0.564 & 0.050 & 0.050 & 0.050\tabularnewline
 &  & 500 & 0.045 & 0.051 & 0.801 & 0.051 & 0.051 & 0.051\tabularnewline
 &  & 1000 & 0.045 & 0.048 & 0.861 & 0.048 & 0.048 & 0.048\tabularnewline
\addlinespace
0.1 & 0 & 100 & $<$0.001 & 0.004 & 0.070 & 0.049 & 0.026 & 0.026\tabularnewline
 &  & 500 & 0.001 & 0.009 & 0.126 & 0.050 & 0.035 & 0.035\tabularnewline
 &  & 1000 & 0.002 & 0.017 & 0.191 & 0.049 & 0.042 & 0.041\tabularnewline
\addlinespace
0.2 & 0 & 100 & 0.001 & 0.009 & 0.116 & 0.053 & 0.035 & 0.035\tabularnewline
 &  & 500 & 0.006 & 0.031 & 0.307 & 0.048 & 0.046 & 0.047\tabularnewline
 &  & 1000 & 0.013 & 0.042 & 0.457 & 0.048 & 0.048 & 0.047\tabularnewline
\addlinespace
0.3 & 0 & 100 & 0.002 & 0.016 & 0.187 & 0.048 & 0.039 & 0.038\tabularnewline
 &  & 500 & 0.016 & 0.046 & 0.488 & 0.048 & 0.048 & 0.049\tabularnewline
 &  & 1000 & 0.030 & 0.050 & 0.626 & 0.050 & 0.050 & 0.050\tabularnewline
\addlinespace
0.4 & 0 & 100 & 0.005 & 0.025 & 0.268 & 0.049 & 0.045 & 0.044\tabularnewline
 &  & 500 & 0.028 & 0.051 & 0.615 & 0.051 & 0.052 & 0.052\tabularnewline
 &  & 1000 & 0.039 & 0.049 & 0.724 & 0.049 & 0.049 & 0.049\tabularnewline
\bottomrule
\end{tabular}

\end{table}

The results for the type I error are shown in Table~\ref{t:typeI}.
The S-test maintains type I error around the nominal level. 
The type I error of the PS-test, ASQ-test, and maxP is lower than that of the S-test 
and approaches the nominal level as $\beta$ or $\gamma$ increases.
The Sobel test is very conservative when $\beta$ or $\gamma$ is zero or very small
and when $n$ is small.
By contrast, the product-normal test is anti-conservative when $\beta$ or $\gamma$ is not 0. 

\setlength\tabcolsep{4pt}
\begin{table}[ht]
  \caption{Empirical Power and Relative Efficiency at the Nominal Significance Level of 0.05}
  \label{t:power}
  \centering
\scalebox{0.79}{
\begin{tabular}{lll>{\raggedleft\arraybackslash}p{1.35cm}>{\raggedleft\arraybackslash}p{1.35cm}>{\raggedleft\arraybackslash}p{1.35cm}>{\raggedleft\arraybackslash}p{1.35cm}>{\raggedleft\arraybackslash}p{1.675cm}>{\raggedleft\arraybackslash}p{1.35cm}>{\raggedleft\arraybackslash}p{1.35cm}>{\raggedleft\arraybackslash}p{1.35cm}>{\raggedleft\arraybackslash}p{1.675cm}}
\toprule
\multicolumn{3}{c}{ } & \multicolumn{5}{c}{Power} & \multicolumn{4}{c}{Relative efficiency} \tabularnewline
\cmidrule(l{3pt}r{3pt}){4-8} \cmidrule(l{3pt}r{3pt}){9-12}
$\beta$ & $\gamma$ & $n$ & Sobel & maxP & S-test & PS-test & ASQ-test & Sobel & S-test & PS-test & ASQ-test\tabularnewline
\midrule
0.05 & 0.03 & 100 & $<$0.001 & 0.003 & 0.052 & 0.028 & 0.026 & 0.04 & 20.02 & 10.65 & 9.85\tabularnewline
 &  & 500 & 0.001 & 0.010 & 0.054 & 0.037 & 0.035 & 0.07 & 5.70 & 3.87 & 3.73\tabularnewline
 &  & 1000 & 0.002 & 0.019 & 0.064 & 0.052 & 0.050 & 0.10 & 3.27 & 2.67 & 2.59\tabularnewline
\addlinespace
0.1 & 0.1 & 100 & 0.001 & 0.011 & 0.053 & 0.039 & 0.038 & 0.10 & 4.69 & 3.44 & 3.38\tabularnewline
 &  & 500 & 0.033 & 0.122 & 0.148 & 0.149 & 0.148 & 0.27 & 1.21 & 1.21 & 1.21\tabularnewline
 &  & 1000 & 0.152 & 0.307 & 0.317 & 0.320 & 0.317 & 0.49 & 1.03 & 1.04 & 1.03\tabularnewline
\addlinespace
 & 0.2 & 100 & 0.007 & 0.037 & 0.065 & 0.061 & 0.061 & 0.18 & 1.73 & 1.64 & 1.64\tabularnewline
 &  & 500 & 0.132 & 0.198 & 0.198 & 0.198 & 0.198 & 0.67 & 1.00 & 1.00 & 1.00\tabularnewline
 &  & 1000 & 0.311 & 0.351 & 0.351 & 0.351 & 0.351 & 0.89 & 1.00 & 1.00 & 1.00\tabularnewline
\addlinespace
 & 0.3 & 100 & 0.023 & 0.066 & 0.075 & 0.075 & 0.074 & 0.35 & 1.14 & 1.14 & 1.13\tabularnewline
 &  & 500 & 0.175 & 0.202 & 0.202 & 0.202 & 0.202 & 0.87 & 1.00 & 1.00 & 1.00\tabularnewline
 &  & 1000 & 0.327 & 0.342 & 0.342 & 0.342 & 0.342 & 0.96 & 1.00 & 1.00 & 1.00\tabularnewline
\addlinespace
 & 0.4 & 100 & 0.041 & 0.076 & 0.078 & 0.078 & 0.078 & 0.54 & 1.02 & 1.02 & 1.02\tabularnewline
 &  & 500 & 0.190 & 0.202 & 0.202 & 0.202 & 0.202 & 0.94 & 1.00 & 1.00 & 1.00\tabularnewline
 &  & 1000 & 0.337 & 0.345 & 0.345 & 0.345 & 0.345 & 0.98 & 1.00 & 1.00 & 1.00\tabularnewline
\addlinespace
0.4 & 0.1 & 100 & 0.020 & 0.083 & 0.112 & 0.111 & 0.109 & 0.24 & 1.36 & 1.35 & 1.32\tabularnewline
 &  & 500 & 0.485 & 0.595 & 0.595 & 0.596 & 0.596 & 0.82 & 1.00 & 1.00 & 1.00\tabularnewline
 &  & 1000 & 0.860 & 0.883 & 0.883 & 0.883 & 0.883 & 0.97 & 1.00 & 1.00 & 1.00\tabularnewline
\addlinespace
 & 0.2 & 100 & 0.093 & 0.242 & 0.269 & 0.274 & 0.270 & 0.38 & 1.11 & 1.13 & 1.11\tabularnewline
 &  & 500 & 0.971 & 0.987 & 0.987 & 0.987 & 0.987 & 0.98 & 1.00 & 1.00 & 1.00\tabularnewline
 &  & 1000 & 1.000 & 1.000 & 1.000 & 1.000 & 1.000 & 1.00 & 1.00 & 1.00 & \vphantom{2}1.00\tabularnewline
\addlinespace
 & 0.3 & 100 & 0.225 & 0.414 & 0.425 & 0.428 & 0.426 & 0.54 & 1.03 & 1.03 & 1.03\tabularnewline
 &  & 500 & 0.992 & 0.994 & 0.994 & 0.994 & 0.994 & 1.00 & 1.00 & 1.00 & 1.00\tabularnewline
 &  & 1000 & 1.000 & 1.000 & 1.000 & 1.000 & 1.000 & 1.00 & 1.00 & 1.00 & \vphantom{1}1.00\tabularnewline
\addlinespace
 & 0.4 & 100 & 0.348 & 0.485 & 0.487 & 0.488 & 0.488 & 0.72 & 1.00 & 1.01 & 1.00\tabularnewline
 &  & 500 & 0.993 & 0.994 & 0.994 & 0.994 & 0.994 & 1.00 & 1.00 & 1.00 & 1.00\tabularnewline
 &  & 1000 & 1.000 & 1.000 & 1.000 & 1.000 & 1.000 & 1.00 & 1.00 & 1.00 & 1.00\tabularnewline
\bottomrule
\end{tabular}
}

Note: Relative efficiency is the power relative to maxP
\end{table}

The results for power are given in Table~\ref{t:power}. 
(The results for the product-normal test are omitted
due to its highly inflated type I error.)
The Sobel test is the least powerful, especially for small $n$ and small effect sizes.
The PS-test and ASQ-test are nearly as powerful as the S-test. In addition, 
they are considerably more powerful than maxP when effect sizes are small. 

\begin{figure}[ht]
   \begin{center}
  \includegraphics[width=4.5in]{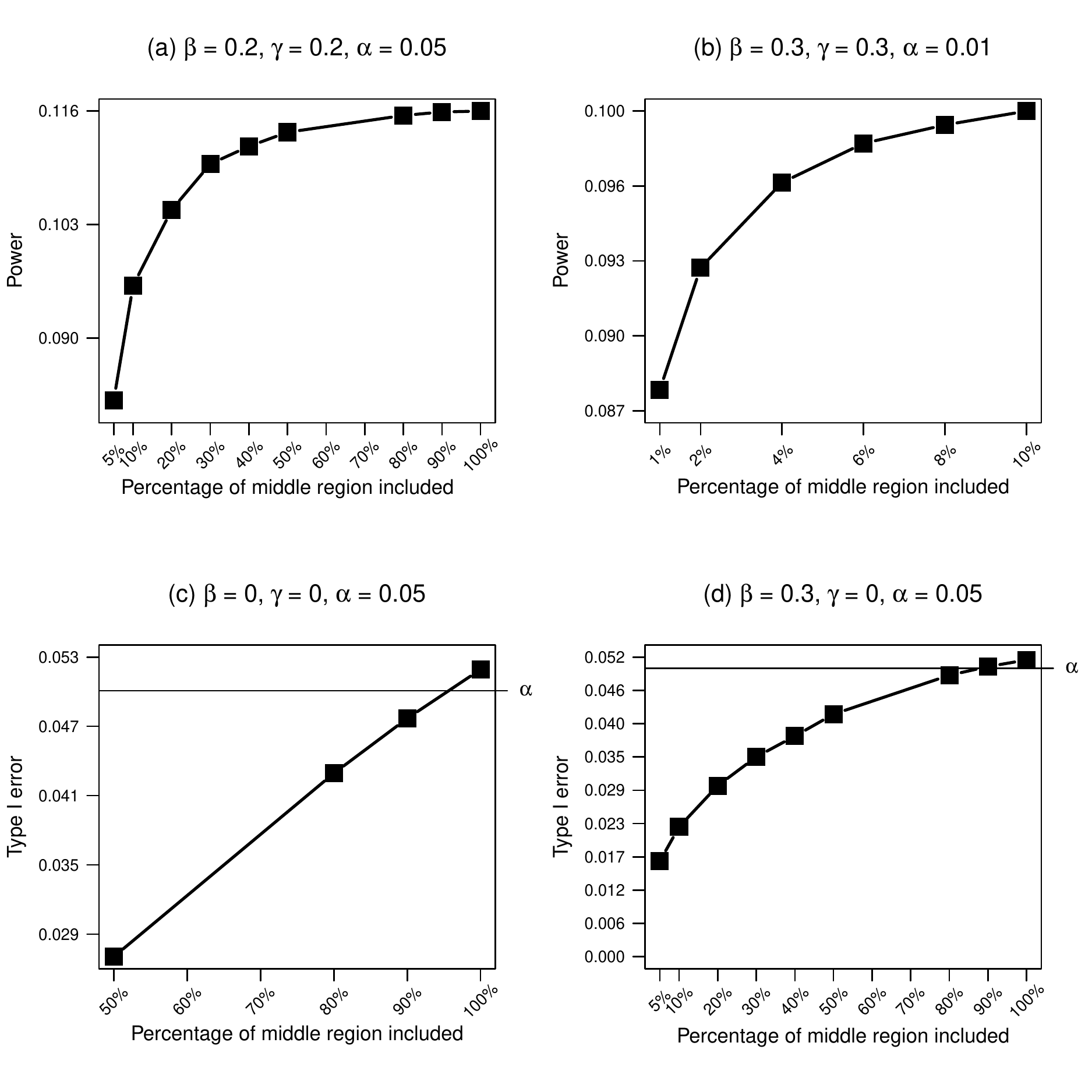}
   \end{center}
   \caption{Power and type I error for the PS-test across different 
              lengths of the center band under 4 scenarios.} \label{figBands}
\end{figure}

Figure~\ref{figBands} shows the changes in the power and type I error when the length 
of the bands for the PS-test varies. 
For small $n$, a larger rejection region greatly increases the power (Figure~\ref{figComb}a) 
but also increases the type I error (Figures~\ref{figComb}c and \ref{figComb}d).
At small $\alpha$, only a small increase in the rejection region is necessary to achieve a large
increase in power (Figure~\ref{figComb}b) without markedly increasing 
the type I error. While different analyses may necessitate different limits, 
the simulation results suggested
limiting the bands of the PS-test to 50\% of the possible extension.
A similar conclusion was reached on the squares of the ASQ-test.

\section{Application to SPIROMICS}
\label{s_results_spiro}  

SPIROMICS is a multi-center study designed to guide future development of therapies
for COPD patients \cite{Couper2014}.
Between November 2011 and January 2015, the study enrolled over 2,900 patients
with varying disease severity. Participants underwent a baseline visit that included a variety of 
measurements, and many different biospecimens were collected and stored.
A major goal of the study was to identify biomarkers as intermediate outcomes in 
order to reliably predict clinical benefits. 

A biomarker panel for 114 blood proteins was assayed through multiple 
Myriad-RBM multiplex technologies. The biomarkers were selected because of known 
or potential links to COPD pathophysiology \cite{ONeal2014, Sun2016}. 
We removed 24 biomarkers with fewer than 500 measurements
and excluded the patients without measurements.
We replaced any measurement below the detection limit by half of the detection limit 
and set any measurement above the detection limit to the upper limit.
Finally, we applied the inverse-normal transformation to each of the remaining 90 biomarkers.

Genotype data for 2,714 participants were obtained from Illumina OmniExpress plus Exome GeneChip,
with a total of 673,688 single nucleotide polymorphisms (SNPs). 
After removing any SNP with greater than 10\% missing values or minor allele frequency less than 1\%,
we were left with 615,535 autosomal SNPs. 

We focused on the phenotype emphysema, which is quantified by the 
percentage of lung voxels greater than or equal to 950 Hounsfield Units on full 
inspiratory CT scans. We considered the 1,589 patients with available phenotype, biomarker, and emphysema data.
We performed principal component analysis on common SNPs and included the top five 
principal components as covariates in the models to account for population stratification.
We also included age, gender, body mass index, smoking pack years, 
and current smoking status as covariates. 

\begin{figure}[ht]
   \begin{center}
  \includegraphics[width=4.5in]{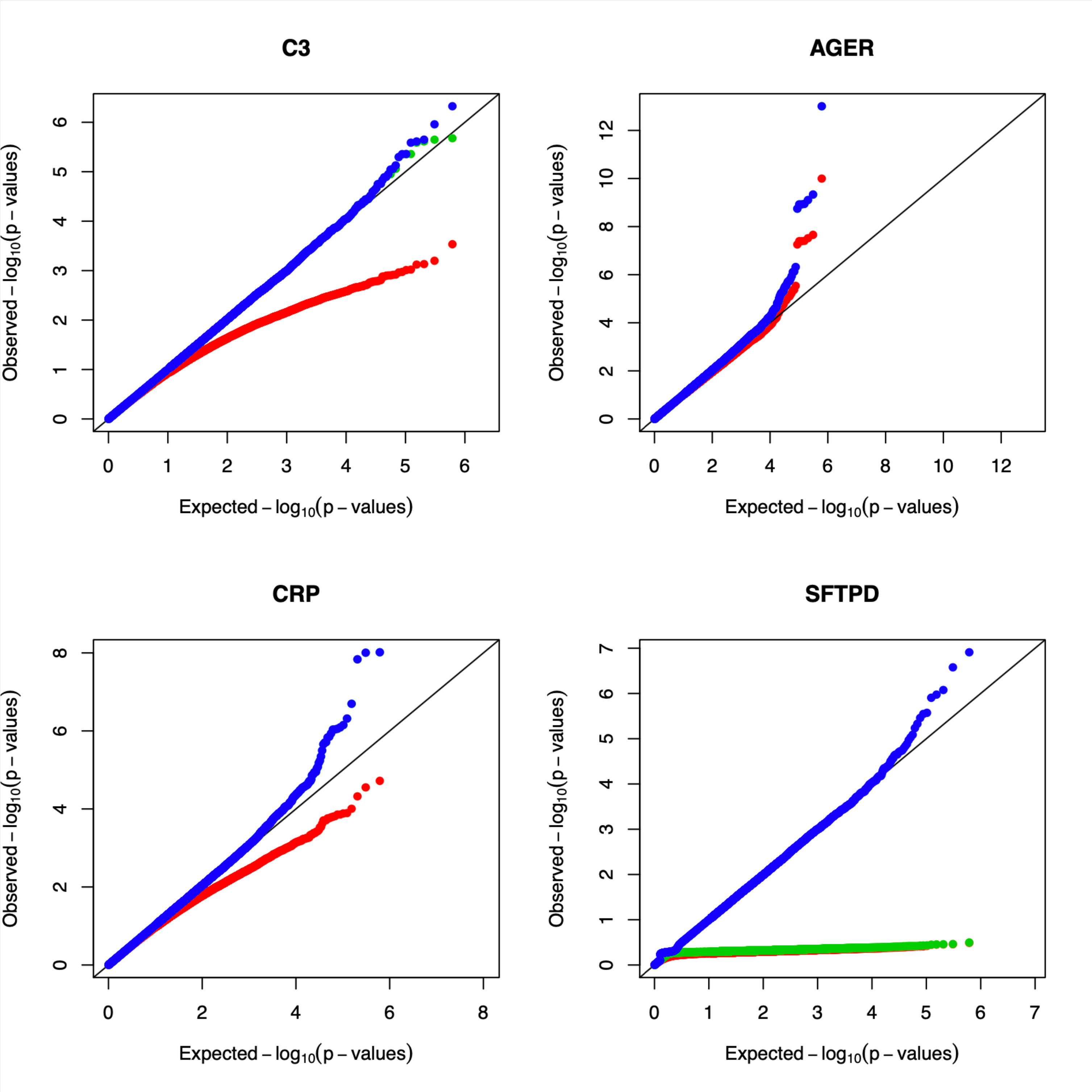}
  \end{center}
   \caption{Quantile-quantile plots of the $\text{-log}_{10}$ $p$-values for testing    
   the indirect effects in the SPIROMICS study: 
   the results for the PS-test, maxP, and Sobel test are 
   shown in blue, green, and red, respectively.
For AGER and CRP, the results of the PS-test and maxP are indistinguishable.                      
For SFTPD, the results of the Sobel test and maxP are indistinguishable. } \label{figqq}
\end{figure}

We conducted mediation analysis for each combination of SNPs and biomarkers, 
with the biomarker as the mediator. Using each SNP as the independent variable, 
we assume there is no unmeasured confounding. 
We tested for indirect effects with the Sobel test, maxP, S-test, 
PS-test, and ASQ-test. 
Figure~\ref{figqq} provides the quantile-quantile (QQ) plots for four biomarkers.
The results for the S-test and ASQ-test are highly similar to those of the 
PS-test and thus are omitted. All four QQ-plots for the PS-test are well behaved. 
The QQ-plots for the Sobel test are highly deflated for three out of the four biomarkers,
and one of the QQ-plots for maxP is also highly deflated.

Using an earlier version of the SPRIROMICS data, Sun et al. \cite{Sun2016} found evidence of indirect
effect for biomarker C3; however, Figure~\ref{figqq} shows no such evidence. 
Unlike Sun et al. \cite{Sun2016}, the PS-test, maxP, 
and Sobel test found an indirect effect through AGER; 
this finding is consistent with the report of Zhang et al. \cite{Zhang2018}. 
In addition, the PS-test and maxP found a potential indirect effect in CRP, 
which is consistent with Aref and Refaat \cite{Aref2014}, whereas the Sobel test did not. 
The PS-test found a potential indirect effect for SFTPD, whereas the Sobel test and maxP did not.
This result, which is consistent with the findings of Obeidat et al. \cite{Obeidat2017}, can further the
understanding of a biological process associated with COPD.

\section{Discussion}
\label{s_discuss}

Existing methods for detecting indirect effects in mediation analysis are either 
overly conservative or anti-conservative.
We have presented powerful tests that preserve the type I error.
Such tests are much needed in the field of genetics, where the effects tend to be small
and controlling the type I error is paramount. 
By making use of the cumulative probabilities from any distribution, 
our methods extend the S-test to allow test statistics that are 
not normally distributed. \cite{Berger1997}.  % change
In addition, we address the inherent limitations of the S-test.

In many genetics studies, such as SPIROMICS, mediators may not be measured on
all study participants because of cost or other constraints.
It is possible to construct an appropriate likelihood to accommodate missing values
on a mediator \cite{Lin2020}. 
The resulting maximum likelihood estimators of $\beta$ and $\gamma$
are generally no longer independent.
We are currently extending our methods to allow dependence of the estimators.

We have focused on the case of a single mediator.
In some applications, investigators are interested in multiple mediators \cite{VanderWeele2014},
which may jointly affect the response variable or may affect one another. 
We are currently extending our framework to such scenarios.

%\backmatter

\section*{ACKNOWLEDGMENTS}

We thank Dr. Chris Shendhal for processing the SPIROMICS data. 
Research reported in this publication was supported by the National Institute Of 
Health under Award Number R01HG009974. The content is solely the responsibility 
of the authors and does not necessarily represent the official views of the 
National Institute of Health.

%\bibliographystyle{unsrtnat}
% \bibliographystyle{vancouver}
%\bibliographystyle{apalike}
%\bibliography{fullBib}

%%%%%%%%%%%%%%%%%%%%%%%%%%%%%%%%%%%%%%%%%%%%%%%%%%%%%%%%%%%%%%%%%%%%%%
%%%%%%%%%%%%%%%%%% Manual Bibliography%%%%%%%%%%%%%%%%%%%%%%%%%%%%%%%%
%%%%%%%%%%%%%%%%%%%%%%%%%%%%%%%%%%%%%%%%%%%%%%%%%%%%%%%%%%%%%%%%%%%%%%

\section*{Appendix: Potential Inflation of the Type I Error for the PS-Test}
\label{appn}
Due to the symmetric nature of the rejection region, the potential inflation of the Type I error 
for the PS-test is the same when $\gamma = 0$ and $\beta \ne 0$ versus when $\beta = 0$ and $\gamma \ne 0$. 
Thus, we assume that $\gamma = 0$ and $\beta \ne 0$, in which case $U_{\gamma}$ is standard uniform. 
The probability that $(U_{\beta}, U_{\gamma})$ lies within the rejection region depends on the value of $\alpha$ 
(just like the original S-test) and on the value of $U_{\beta}$ (unlike the original S-test). 
Let $g(u_{\beta})$ denote the probability that $H_0$ is rejected for the chosen $\alpha$ 
when $U_{\beta}=u_{\beta}$, and let $f_{U_{\beta}}(u_\beta)$ denote the density function of $U_{\beta}$. 
The probability of making a type I error at the significance 
level $\alpha$ equals $\int_0^1 g(x) f_{U_{\beta}}(x) dx$. We can determine the noncentrality parameter
of $T_{\beta}$ that causes the largest type I error for any value of $\alpha$ and then determine
the maximum inflation of the type I error over all possible values of $\alpha$.

We use numerical integration to calculate the type I error. 
We consider both small-sample and asymptotic scenarios, 
using a noncentral $t$-distribution with five degrees of freedom and a normal distribution 
with mean equal to the noncentrality parameter and unit variance. In the small-sample scenario, 
the maximum possible type I error rate occurs when 
$\alpha \approx 0.002$ and is approximately $1.0001$ times $\alpha$. 
In the asymptotic case, the maximum type I error occurs when $\alpha \approx 0.028$
and is approximately $1.0084$ times $\alpha$. In each case, the increase 
in the type I error for the PS-test is less than 1\% of $\alpha$, and 
more common choices of $\alpha$ have even lower inflation of the type I error.

% \label{lastpage}

\end{document}